\documentclass[lineno]{jfm}
\usepackage{graphicx}
\usepackage{newtxtext}
\usepackage{newtxmath}
\usepackage{amssymb}
\usepackage{natbib}
\usepackage{hyperref}
\hypersetup{
    colorlinks = true,
    urlcolor   = blue,
    citecolor  = blue,
    linkcolor  = black,
}

\newcommand{\RomanNumeralCaps}[1]
\linenumbers
\usepackage[acronym]{glossaries}

\title{On the ventilation of surface-piercing hydrofoils under steady-state conditions}

\author{Manuel Aguiar Ferreira\corresp{\email{M.Ferreira@tudelft.nl}},
        Carlos Navas Rodríguez,
        Gunnar Jacobi,
        Daniele Fiscaletti,
        Arnoud Greidanus,
        \and Jerry Westerweel
        }

\affiliation{Faculty of Mechanical Engineering, Delft University of Technology, Delft, Netherlands}

\begin{document}
\maketitle

\begin{abstract}
The present study experimentally investigates the onset of ventilation of surface-piercing hydrofoils.
Under steady-state conditions, the depth-based Froude number $Fr$ and the angle of attack $\alpha$ define regions where distinct flow regimes are either locally or globally stable.
To map the boundary between these stability regions, the parameter space $(\alpha,Fr)$ was systematically surveyed by increasing $\alpha$ until the onset of ventilation, while maintaining a constant $Fr$.
Two simplified model hydrofoils were examined: a semi-ogive profile with a blunt trailing edge and a modified NACA 0010-34.
Tests were conducted in a towing tank under quasi-steady-state conditions for aspect ratios of $1.0$ and $1.5$, and $Fr$ ranging from $0.5$ to $2.5$.
Ventilation occurred spontaneously for all test conditions as $\alpha$ increased.
Three distinct trigger mechanisms were identified: nose, tail, and base ventilation.
Nose ventilation is prevalent at $Fr<1.0$ and $Fr<1.25$ for aspect ratios of $1.0$ and $1.5$, respectively, and is associated with an increase in the inception angle of attack.
Tail ventilation becomes prevalent at higher $Fr$, where the inception angle of attack takes a negative trend.
Base ventilation was observed only for the semi-ogive profile but did not lead to the development of a stable ventilated cavity.
Notably, the measurements indicate that the boundary between bistable and globally stable regions is not uniform and extends to significantly higher $\alpha$ than previously estimated.
A revised stability map is proposed to reconcile previously published and current data, demonstrating how two alternative paths to a steady-state condition can lead to different flow regimes.
\end{abstract}

\begin{keywords}
Authors should not enter keywords on the manuscript, as these must be chosen by the author during the online submission process and will then be added during the typesetting process (see \href{https://www.cambridge.org/core/journals/journal-of-fluid-mechanics/information/list-of-keywords}{Keyword PDF} for the full list).  Other classifications will be added at the same time.
\end{keywords}

\vfill
\pagebreak 

\newacronym{dut}{DUT}{Delft University of Technology}
\newacronym{fw}{FW}{\emph{fully wetted}}
\newacronym{fv}{FV}{\emph{fully ventilated}}
\newacronym{lsb}{LSB}{laminar separation bubble}
\newacronym{ptu}{PTU}{programmable timing unit}
\newacronym{pv}{PV}{\emph{partially ventilated}}
\newacronym{rans}{RANS}{Reynolds-averaged Navier-Stokes}

\section{Introduction}

Hydrofoils have become increasingly popular in recent decades for application to fast passenger ferries and high-performance sailing.
They lift the hull up and out of the water as speed increases, reducing the wetted surface area.
In turn, the combined frictional, pressure and wave-making resistance significantly decreases, improving the overall sailing efficiency and performance compared to equivalent bare-hull designs.
Hydrofoils are broadly classified as surface piercing, whose lifting surface rises above the waterline, or fully submerged.
Surface-piercing hydrofoils are inherently stable and do not require additional lift-control systems for safe operation. 
However, interacting with the free surface makes them prone to ventilation, causing a sudden drop in lift and flutter, with potentially catastrophic consequences.

\citet{Wadlin1958} identified two necessary conditions for the onset of ventilation: firstly, the development of a separated flow region at subatmospheric pressure and, secondly, the establishment of a steady supply of non-condensible gas or water vapour into that region.
Surface-piercing hydrofoils operating at shallow depths, with an aspect ratio of $\mathcal{O}(1)$, are particularly subject to air ingress from the atmosphere through the free surface.
This phenomenon is known as \emph{natural ventilation} \citep{Acosta1973}.
Air is either drawn locally from a narrow region adjacent to the low-pressure side of the hydrofoil or remotely through the core of the tip vortex.
The air entrainment process depends on the prevalent trigger mechanism, whereby the air first breaches the free surface, forming a stable ventilated cavity.
Early studies, including those by \citet{Kiceniuk1954}, \citet{Wetzel1957}, \citet{Ramsen1957}, \citet{Breslin1959}, \citet{Waid1968}, \citet{Rothblum1969}, and \citet{Wright1972}, have identified and documented these trigger mechanisms.
Their nature is diverse and seemingly stochastic, posing a significant challenge in accurately predicting the onset of natural ventilation.

The flow around surface-piercing hydrofoils is classified into three regimes based on the stability and topology of the ventilated cavity: \gls{fw}, \gls{pv}, or \gls{fv}, as formally defined by \citet{Harwood2016}.
In the fully wetted regime, air entrainment is not sustained or is confined to the wake region of hydrofoils with a blunt trailing edge.
Flow separation may occur, particularly at high angles of attack, but a steady supply of air has not yet been established --- only one of two necessary conditions for the onset of ventilation is satisfied.
The partially ventilated regime features a stable or unstable cavity over the suction side of the hydrofoil, which extends along the chord up to the trailing edge.
The cavity is not uniform along the span. 
It tends to be broader at the root (near the free surface) and taper off towards the tip owing to three-dimensional effects.
Finally, the ventilated cavity stabilises, reaches the tip of the hydrofoil, and may extend several chord lengths beyond the trailing edge in the fully ventilated regime.

Under standard atmospheric pressure and steady-state conditions, the relevant parameters governing the development and elimination processes of ventilation, as well as the transition between flow regimes, are the geometric angle of attack $\alpha$, the forward speed $u$, and the static immersion depth of the hydrofoil $h$ \citep{Swales1974, Young2017}. 
The last two parameters define the depth-based Froude number, expressed as
\begin{equation}
    Fr = \frac{u}{\sqrt{g\,h}},
\end{equation}
where $g$ is the gravitational acceleration.
A significant effort has been made to map the flow regimes as a function of the angle of attack and the depth-based Froude number.
\citet{Fridsma1963} identified regions where each flow regime is globally stable and interfacial regions where two regimes are locally stable, which \citet{Harwood2016} later termed \emph{bistable}. 
Within these bistable regions, the flow exhibits a hysteresis behaviour, first reported by \citet{Kiceniuk1954}, \citet{Wetzel1957}, and \citet{Breslin1959}. 
Once the flow transitions from the \gls{fw} regime to the \gls{pv} or \gls{fv} regime, it remains ventilated unless either one or both parameters are adjusted to restore the original globally stable conditions.
Subsequent studies by \citet{Waid1968}, \citet{Rothblum1969}, \citet{Swales1974}, and \citet{Heckel1974}, amongst others, have produced stability maps, albeit incomplete or lacking sufficient resolution.
Incidentally, variations in the geometry of the model hydrofoil and test conditions (\citealp[see][]{Young2017} for an overview) make it difficult to piece together observations from various sources.
The most comprehensive stability map was only recently obtained following an extensive experimental campaign by \citet{Harwood2016}.
The stability map is not universal but specific to the model hydrofoil they tested at a given aspect ratio.
Nevertheless, it remains a valuable tool for predicting ventilation under steady-state conditions, and it provides a basis for investigating more realistic scenarios, accounting for the effect of waves and fluid-structure interaction \citep{Young2022, Young2023a, Young2023b}.

The past decade has seen a growing interest in numerical methods for modelling natural ventilation of surface-piercing hydrofoils.
\gls{rans} simulations performed by \citet{Harwood2014}, \citet{Matveev2018}, \citet{Charlou2019}, \citet{Matveev2019}, \citet{Andrun2021}, and \citet{Zhi2022} have successfully reproduced measurements from \citet{Harwood2016}.
These simulations offer valuable insights into the complex behaviour of ventilation by providing a more detailed description of the flow field.
However, when numerical models disagree or do not meet original assumptions, they require validation against a reference data set. 
Thus, research continues to lean heavily on experiments conducted in flume and towing tank facilities, which are not without limitations.
Small floating debris and residual surface perturbations, carried over between consecutive runs or caused by model vibrations, can prematurely trigger ventilation, contributing to increased measurement variability.
Careful design of experiments can mitigate these effects, but other factors associated with the experimental procedure may play a role.

In particular, mapping the boundary between stability regions involves a systematic survey of the parameter space $(\alpha, Fr)$, where the model hydrofoil is accelerated to a target forward speed or Froude number while maintaining a fixed angle of attack \citep[e.g.][]{Breslin1959, Harwood2016}.
Ventilation occurs spontaneously during the initial acceleration phase or is subsequently triggered artificially to determine whether the flow lies in a globally stable or a bistable region.
Artificial triggers disturb the free surface near the leading edge, for example, using thin wires or pressurised air jets.
\citet{Rothblum1969} introduced an alternative approach, recently adopted by \citet{Charlou2019}, where the angle of attack is increased until the onset of ventilation while maintaining a constant Froude number.
Under suitable quasi-steady-state conditions, it is generally assumed that the stability map is independent of the surveying procedure.
This assumption underpins most studies on the ventilation of surface-piercing hydrofoils, yet it remains largely untested.
\gls{rans} simulations performed by \citet{Charlou2019} suggest that the inception angle of attack, at which the air first breaches the free surface, is considerably higher following the alternative approach.
They attributed this discrepancy to transient effects, offering no insights into other potential factors.

The present study examines this critical assumption by surveying the parameter space under quasi-steady conditions.
The experiment primarily focuses on the alternative approach \citep{Rothblum1969} described above, for which data is lacking, but also the conventional approach for validation purposes. 
It considers two simplified models of a surface-piercing hydrofoil.
As outlined in \S \ref{sec: methods}, one features a semi-ogive profile with a blunt trailing edge, introduced by \citet{Harwood2016}, and the other a modified NACA 0010-34.
The results, presented in \S \ref{sec: results}, include a detailed analysis of the trigger mechanisms of ventilation, the inception angle of attack, and the hydrodynamic coefficients.
The flow stability map is discussed in depth in \S \ref{sec: discussion}, and the main findings are summarised in \S \ref{sec: conclusion}.

\section{Experimental methods}
\label{sec: methods}
Experiments were conducted in the towing tank of the Ship Hydromechanics Laboratory at \gls{dut}.
The tank is $142.0\,$m long by $4.2\,$m wide, and the water depth was set to $2.2\,$m.
A towing carriage runs along two parallel rails installed on either side, reaching a top speed of $7\,$m.s$^{-1}$ and an acceleration of $0.9\,$m.s$^{-2}$.
The design of the model hydrofoils is presented in \S \ref{subsec: model}, followed by an outline of the test conditions in \S \ref{subsec: tests}.
A criterion for the quasi-steady-state condition is established in \S \ref{subsec: quasi-steady}, which prescribes the rate of change of the angle of attack.
The measurement instruments and calibration procedures are described in \S \ref{subsec: force_balance} through to \S  \ref{subsec: alignment}.
The coordinate system follows the convention that $x$, $y$, and $z$ are the longitudinal, transversal, and vertical axes, and $\gamma$, $\beta$, and $\alpha$ represent the Euler angles around those axes.

\subsection{Model hydrofoils}
\label{subsec: model}
\begin{figure}
    \centering
    \includegraphics{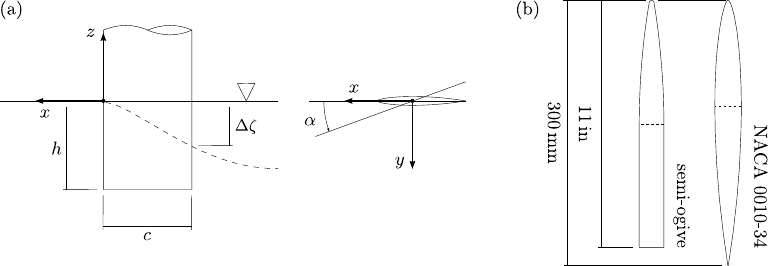}
    \caption{(a) Schematic of a surface-piercing hydrofoil, illustrating key geometric parameters: chord-length $c$, static-immersion depth $h$, free surface deformation at the trailing edge $\Delta \zeta$, and angle of attack $\alpha$.
    (b) Sectional profiles of the semi-ogive (left) and the NACA 0010-34 (right).}
    \label{fig: model}
\end{figure}

\begin{figure}
    \centering
    \includegraphics{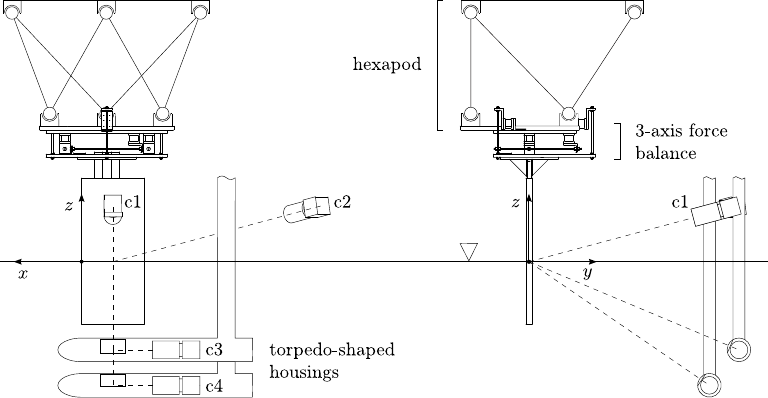}
    \caption{Illustration of the experimental setup.
    The model is securely fastened to a three-axis force balance, supported by a hexapod with electric linear actuators. 
    A pair of digital cameras, c$1$ and c$2$, is arranged in a stereo configuration above the waterline, and another pair, c$3$ and c$4$, is submerged underwater inside watertight torpedo-shaped housings.
    The relative position of the cameras to the model is not to scale.}
    \label{fig: setup}
\end{figure}

Two simplified models of a surface-piercing hydrofoil, represented in figure \ref{fig: model}, were examined.
Each model consists of a streamlined vertical strut with a uniform cross-section along the span of chord length $c$, set at a static immersion depth $h$, and an arbitrary geometric angle of attack $\alpha$.
The cross-sectional profiles, semi-ogive and NACA 0010-34, are relatively thin with a sharp leading edge to encourage flow separation at moderate angles of attack $\alpha \lesssim 10$\,$^{\circ}$.
The semi-ogive features a circular-blunted, tangent-ogive forward section transitioning to a square section with a blunt trailing edge.
Its dimensions are the same as the model hydrofoil of \citet{Harwood2016}.
The modified NACA 0010-34 has a reduced leading-edge radius set to $1/4$ of the standard four-digit profile, and the position of the maximum thickness is farther aft at $40\,$\% of the chord.

The models were $3$D printed in a photopolymer resin as a single piece with a gyroid infill and a wall thickness of $3\,$mm.
The spars consist of three $40\times20\times3\,$mm$^3$ rectangular hollow section steel beams bonded in place with epoxy resin, providing most of the bending and torsional stiffness.
The surfaces were sanded before receiving a coat of primer, followed by two coats of marine-grade polyurethane paint, ensuring a smooth, matt finish.
Square grids with a pitch of $c/14$ and $c/15$ were drawn with a marker on the suction side of the semi-ogive and the NACA0010-34, respectively.
As illustrated in figure \ref{fig: setup}, the models were securely fastened to a three-axis force balance, which, in turn, was mounted to a hexapod with electrical linear actuators that can be easily reconfigured and execute prescribed motion profiles.

\subsection{Calm-water tests}
\label{subsec: tests}
\begin{figure}
    \centering
    \includegraphics{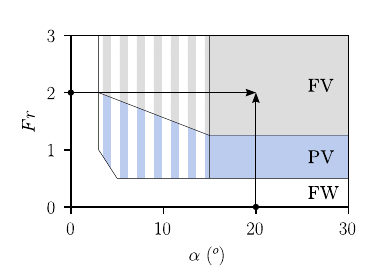}
    \includegraphics{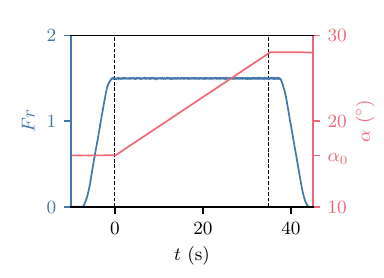}
    \vspace{-10pt}
    \caption{(a) Simplified flow stability map for the semi-ogive hydrofoil at an aspect ratio of $1$, adapted from \citet{Harwood2016}, with the depth-based Froude number $Fr$ along the vertical axis and the angle of attack $\alpha$ along the horizontal axis.
    Solid colours represent the regions of global stability of the fully wetted (FW), partially ventilated (PV), and fully ventilated (FV) regimes, while striped patterns represent bistable regions.
    Arrows illustrate two alternative approaches to exploring the parameter space.
    (b) Time series of the parameters for an arbitrary test case as they vary along the $\alpha-\text{axis}$.}
    \label{fig: stability_map_1}
\end{figure}

A simplified version of the flow stability map produced by \citet{Harwood2016} for the semi-ogive at an aspect ratio of $1.0$ is presented in figure \ref{fig: stability_map_1}a, with the depth-based Froude number and the angle of attack along the vertical and horizontal axes, respectively.
The map features three global stability regions associated with each flow regime: \gls{fw} at low Froude numbers and/or low angles of attack (white-shaded area), \gls{pv} at moderate Froude numbers and high angles of attack (blue-shaded area), and \gls{fv} at high Froude numbers and angles of attack (grey-shaded area).
There are also interfacial regions where the flow exhibits a bistable behaviour, assuming either of two locally stable regimes.
Specifically, the \gls{fw} regime shares bistable regions at angles of attack below $15\,^{\circ}$ both with the \gls{pv} regime at moderate Froude numbers (white and blue striped area) and with the \gls{fv} regime at high Froude numbers (white and grey striped area).

The conventional approach to map the boundary between stability regions involves surveying the parameter space along the $Fr-\text{axis}$ \citep{Harwood2016}, as indicated by the vertical arrow in figure \ref{fig: stability_map_1}a.
Accordingly, the model hydrofoil is set to an arbitrary angle of attack and accelerated to a target forward speed or Froude number under quasi-steady-state conditions.
For angles of attack above $15\,^{\circ}$, the flow is expected to transition spontaneously from the \gls{fw} regime to the \gls{pv} regime at a Froude number of approximately $0.5$ and then again to the \gls{fv} regime at about $1.25$.
For angles of attack below $15\,^{\circ}$, as the \gls{fw} regime shares bistable regions with the \gls{pv} and \gls{fv} regimes, the flow does not become ventilated unless the transition is triggered artificially.
Repeating this process systematically across the full range of angles of attack, identifying the point at which the flow transitions between regimes, draws a complete picture of the stability map.

The present study adopted an alternative approach by surveying the parameter space along the $\alpha-\text{axis}$, as shown in figures \ref{fig: stability_map_1}a-b.
The hydrofoil is initially set at a shallow angle of attack $\alpha_0<15\,^{\circ}$, at which spontaneous ventilation is not anticipated, and accelerated at $0.5\,$m.s$^{-2}$ to a target forward speed or Froude number.
The angle of attack is then gradually increased at a constant rate until the onset of ventilation.
The rate of change of the angle of attack $\dot\alpha$ is sufficiently slow to ensure quasi-steady-state conditions, minimising surface disturbances and the effect of unsteady hydrodynamics, based on the empirical criterion derived in \S\ref{subsec: quasi-steady}.

Tests were performed for aspect ratios of $1$ and $1.5$, with Froude numbers ranging from $0.5$ to $2.5$ in increments of $0.25$, limited by the full scale of the force balance and the loading capacity of the hexapod.
The free-surface elevation was measured between consecutive runs using a resistive wave probe mounted alongside the model to ensure nominal calm-water conditions.
The waiting period was sufficiently long for the root-mean-square of the free-surface elevation to drop below $1\,$ mm ($\sim0.35\,$\% of the chord length).
Each test case was repeated at least $5$ times for the semi-ogive and $3$ times for the NACA 0010-34 to assess repeatability.
The total number of runs was $150$.
Additional tests were performed following the conventional approach for validation at an aspect ratio of $1$ and a Froude number of $1.5$.
They aimed to confirm that previously published data can be reproduced and to assess whether the quasi-steady-state assumption is well approximated.

\begin{table}
    \begin{center}
        \def~{\hphantom{0}}
        \def\arraystretch{1.2}
        \begin{tabular}{c c c}
            $Fr$ & $AR$ & $\Delta\tau$ $(/^\circ)$ \\
            $[0.50:0.25:2.50]$ & $1.0$ & $30$ \\
            $[0.50:0.25:1.75]$ & $1.5$ & $20$
        \end{tabular}
        \caption{Overview of the calm-water tests. The parameters are the depth-based Froude number $Fr$, the aspect ratio $AR$, and the rate of change of the angle of attack as the number of convective timescales per degree $\Delta\tau$, defined in \S\ref{subsec: quasi-steady}. }
        \label{tab: test_conditions}
    \end{center}
\end{table}

\subsection{Quasi-steady-state criterion}
\label{subsec: quasi-steady}

\begin{figure}
    \centering
    \includegraphics{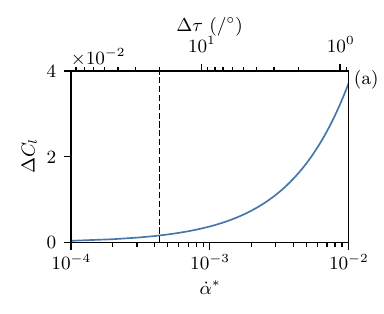}
    \includegraphics{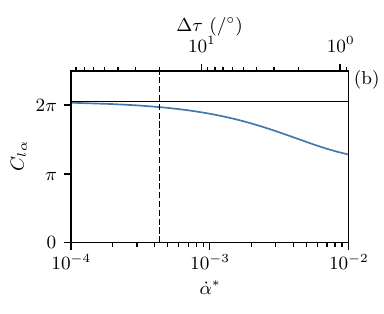}
    \vspace{-10pt}
    \caption{Effect of the rate of change of angle of attack on the lift offset $\Delta C_l$  (a) and the lift-curve slope ${C_l}_{\alpha}$ (b) for a symmetric profile with a thickness ratio $t/c$ of $0.1$ rotating about the mid-chord. 
    The rate of change of the angle of attack is given in non-dimensional form $\dot\alpha^* = c \dot\alpha / (2 u)$ on the bottom horizontal axis and as the number of convective timescales per degree $\Delta\tau = (1 / \dot\alpha) (u/c)$ on the top horizontal axis (reversed), respectively.
    The solid-black line indicates the steady-state lift-curve slope and the dashed-black line $\Delta\tau = 20$.}
    \label{fig: potential_flow}
\end{figure}

Despite numerous studies on the ventilation of surface-piercing hydrofoils, it remains unclear what a meaningful criterion for the quasi-steady state could be.
\citet{Harwood2016} defined the quasi-steady state as \emph{well approximated} at any given time and period throughout the acceleration phase when the flow exhibits no significant topological change, and the towing velocity remains within $\pm3\,$\%.
They additionally showed through dimensional analysis of the hydrodynamic loads that the ratio of the unsteady component due to the added mass to the steady component becomes negligible for Froude numbers larger than $1$, at a rate of acceleration comparable to $0.5\,$m.s$^{-2}$.
This approximation allowed multiple measurements of the hydrodynamic coefficients to be taken at a fixed angle of attack and varying Froude numbers in a single run.

Following a similar argument, the present study assumes the quasi-steady state to be well approximated when the yawing motion has a negligible effect on the hydrodynamic loads, particularly the lift force.
This effect is related to the problem of dynamic stall investigated by \citet{Tupper1983} using potential flow theory applied to a generic Joukowski profile.
He derived expressions that quantify the effect of $\dot\alpha$ on the lift force as a function of the non-dimensional rate of change of the angle of attack $\dot\alpha^* = c \dot\alpha / (2 u)$, camber, thickness ratio, and location of the pivot point. 
Supported by experimental evidence from \citet{Jumper1987}, his model predicts that the hydrofoil experiences an offset in lift coefficient $\Delta C_l$ at rotation onset and the lift-curve slope ${C_l}_{\alpha}$ becomes shallower.
Specifically, for a symmetric profile with a thickness ratio $l/c$ rotating about the half chord, 
\begin{equation}
    \label{eq: Delta_Cl}
    \Delta C_l(\dot\alpha^*, l/c) \, = \, 3.47 \, \dot\alpha^* \left[ 1 + \frac{2}{3} \left(\frac{l}{c}\right)\right]
\end{equation}
and
\begin{equation}
    \label{eq: Cl_alpha}
    {C_l}_{\alpha}(\dot\alpha^*, l/c) \, = \, 3.6 +  2.68 \, \exp\left( -\frac{\dot\alpha^* \times 10^3}{4.216}\right) + \left(\frac{l}{c}\right)^{3/4}.
\end{equation}
Equations \ref{eq: Delta_Cl} and \ref{eq: Cl_alpha} are plotted in figure \ref{fig: potential_flow} as a function of $\dot\alpha^*$ on the bottom horizontal axis and the number of convective timescales per degree $\Delta\tau = (1 / \dot\alpha) (u/c)$ on the top horizontal axis, where $\tau = t (u/c)$.
The results show that for $\Delta\tau > 20$ or $\dot\alpha^* < 4.36\times 10^{-4}$, indicated by the vertical-dashed line, the predicted lift offset is lower than $1.62 \times 10 ^{-3}$ and the relative decrease in the lift-curve slope is less than $4.10\times10 ^{-2}$, suggesting a negligible effect of unsteady hydrodynamics.
The calm-water tests are thus prescribed by $20 < \Delta\tau < 30$ to ensure similar quasi-steady-state conditions across the range of Froude numbers.
This criterion is empirically confirmed a posteriori in \S \ref{subsec: quasi-steady state?} by showing a collapse between measurements obtained under steady-state and quasi-steady-state conditions.

\subsection{Force balance}
\label{subsec: force_balance}
A three-axis force balance, illustrated in figure \ref{fig: setup}, measured the forces in the horizontal plane and the yawing moment.
It comprises two parallel plates connected by a system of linkages and strain gauge load cells, which mechanically decouple the hydrodynamic load components.
The rated capacity of the force balance depends on the loading combination, determined by the towing velocity and the angle of attack.
Along each independent axis, the rated values are $F_{x'} = 400\,$N, $F_{y'} = 1600\,$N, and $M_{z'} = 320\,$N.m.
The load cells were connected to amplifier modules (PICAS from Peekel Instruments), whose analogue outputs were converted into 16-bit digital signals via a multi-channel data logger developed in-house.

The force balance was calibrated off-site using a set of M3-class weights and a pulley arrangement aligned with its reference frame.
The calibration process involved recording the response of the load cells under $30$ different loading conditions.
This process was repeated $3$ times to assess the repeatability error.
Each data point was sampled at a frequency of $1000\,$Hz over $60\,$s to average out noise.
The loading conditions included independent positive and negative forces applied in both the longitudinal and transversal directions and combinations of these forces with a moment applied around the vertical axis.
Forces and moments were computed using the value for the gravitational acceleration $g = 9.81\,\text{m.s}^{-2}$.
A multivariate linear regression model was fitted through the data using the ordinary least squares (OLS) method, which revealed a negligible coupling between load cells, below $0.5\,$\,\%.
The variance-covariance matrix of the parameters was used to evaluate the uncertainty in the forces and the moment associated with the calibration process.
For a confidence interval of $95\,$\%, it is typically lower than $0.1\,$\% and $0.5\,$\%, respectively.

\subsection{Imaging system}
\label{subsec: imaging_system}
A set of cameras secured to the carriage system was used to capture the deformation of the free surface along the chord and the flow over the suction side of the hydrofoil.
As illustrated in figure \ref{fig: setup}, a pair of cameras 2.3 MP Basler acA1920-150um, equipped with 25 mm (focal-length) lenses, were mounted above the waterline.
They were positioned at the same height, one transversely aligned with the model and the other farther downstream.
Another pair of cameras $4\,$MP LaVision Imager MX was mounted below the waterline inside watertight, torpedo-shaped housings.
These cameras were equipped with $28\,$mm lenses, a motorised Scheimpflug adapter, and a remote focus and aperture-control module.
The torpedo-shaped housings feature a water-filled mirror section, which can be rotated around the longitudinal direction to adjust the field of view.
They were aligned transversely with the model and placed half a meter deep to prevent disturbing the free surface.
Fairings were added around the vertical struts to reduce possible hydrodynamic interference.
Illumination above the waterline was provided by two $36\,$W LED panels, mounted perpendicularly to each other, and underwater by submersible LED strips with a combined power of $36\,$W, mounted on the fairing farthest away from the model.
All light sources were equipped with light-diffusing film to ensure uniform illumination and minimise reflections.
A programmable timing unit (LaVision PTU 11) triggered the four cameras simultaneously at a rate of $100\,$Hz, and the images from the Basler and LaVision cameras, which have incompatible formats, were processed by separate systems.
The multi-channel data logger sampled this trigger signal at a rate of $1000\,$Hz synchronously with the force measurements.

The camera pairs were calibrated individually using a 3D calibration plate measuring \(320 \times 320\,\text{mm}^2\), which was temporarily mounted to the hexapod. 
This plate was positioned within the field of view, parallel to the model, at a small stand-off distance. 
Snapshots of the calibration plate were processed to estimate the 2D coordinates of the markers on the image plane.
The intrinsic parameters (focal length, principal point, and distortion coefficients) and the extrinsic parameters (rotation and translation vectors) for each camera were obtained by fitting a pinhole model to these points.
Finally, triangulation using the intrinsic and extrinsic parameters can be performed to determine the coordinates of any point in space.

\subsection{Alignment}
\label{subsec: alignment}
The measurements were taken relative to the moving reference frame of the carriage system $C$, which is aligned with the inertial reference frame of the basin.
Outlined below, a series of steps were taken to ensure the correct alignment of the different parts, namely the model hydrofoil, the force balance, the hexapod, and a 3D optical tracking system (Optotrak Certus NDI 2008), which monitored the position of the working plate of the hexapod (NOTUS V) with submillimeter accuracy.
\setlength{\parskip}{0pt}
\begin{enumerate}
    \item Firstly, the reference frame of the tracking system $T$ was aligned with $C$ by referencing a set of target markers arranged on the free surface in a designated configuration.
    The target markers were later removed.
    \item The reference frame of the hexapod $H$ was aligned with $T$.
    The working plate was displaced along orthogonal directions by $0.5\,$m, and the coordinates of the reference points collected by the tracking system were used to estimate the rotation matrix between the two reference frames.
    The rotation matrix was subsequently applied to the motion instructions of the hexapod.
    The corrected alignment was within $\pm0.05\,^\circ$ around each direction.
    \item The force balance was mounted onto the working plate of the hexapod, following an off-site calibration detailed in \S \ref{subsec: force_balance}.
    The misalignment between the reference frame of the force balance $F$ and $H$ along the vertical direction was measured using a digital inclinometer (Laserliner 081.265A) with a resolution of $0.01\,^\circ$ and an accuracy of $\pm0.05\,^\circ$.
    A $1$-m-long ruler was then mounted longitudinally on the force balance.
    The yaw misalignment was quantified by taking the transversal distance from the ruler to a reference point on the towing carriage as the working plate was displaced longitudinally.
    \item Lastly, the hydrofoil was securely fastened to the force balance.
    The misalignment along the vertical direction was measured using the digital inclinometer.
    The yaw misalignment was estimated by towing the hydrofoil along the basin at $1\,$m.s$^{-1}$ across a narrow range of angles of attack $-5\,^\circ<\alpha<5\,^\circ$ in increments of $1\,^\circ$.
    Drag and lift forces were computed by pre-multiplying the measured force components by the corresponding rotation matrix $R^F$.
    The data points were interpolated to estimate the angle at which the lift force drops to zero.
    The working plate was rotated accordingly to offset any misalignment of the hydrofoil, updating $R^F$ for subsequent measurements.
\end{enumerate}

\section{Results}
\label{sec: results}
This section explores the hydrodynamic behaviour of the model hydrofoils, the semi-ogive and the NACA 0010-34.
Unless otherwise stated, the experiments were conducted by varying the parameters along the $\alpha-\text{axis}$.
The model hydrofoil was accelerated to a target Froude number at a fixed angle of attack, which was then steadily increased until the onset of ventilation under the quasi-steady-state assumption outlined in \S \ref{subsec: quasi-steady}.
A detailed description of the trigger mechanisms of ventilation, whereby the air breaches the free surface, is provided in \S \ref{subsec: trigger_mechanisms}.
Measurements of the inception angle of attack as a function of the depth-based Froude number and the chord-based Reynolds number are presented in \S \ref{subsec: onset_ventilation}.
These measurements are examined considering the prevalent trigger mechanism and transition between flow regimes, providing insights into the physical processes driving ventilation.
Finally, measurements of the hydrodynamic coefficients under steady-state and quasi-steady-state conditions are presented in \S  \ref{subsec: hydrodynamic_coefficients}.

\subsection{Trigger mechanisms}
\label{subsec: trigger_mechanisms}

Three distinct trigger mechanisms were identified: \emph{nose ventilation}, \emph{tail ventilation}, and \emph{base ventilation}.
Although previously documented, they are re-examined here for completeness, leveraging the improved temporal and spatial resolutions of the present data set.
The analysis focuses on time-lapse photography, capturing the free-surface deformation and the development of the ventilated cavity, and incorporates established interpretations of the preceding flow conditions.
The corresponding video recordings are available as supplementary material.

\begin{figure}
    \centering
    \includegraphics{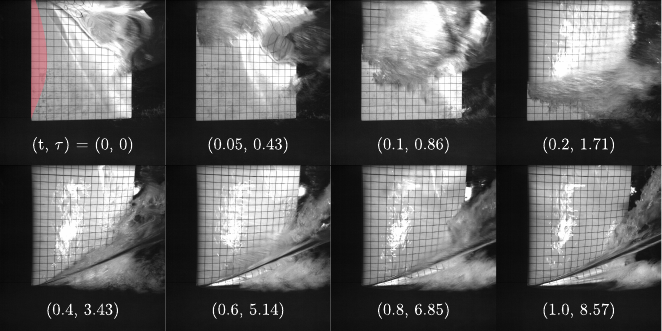}
    \caption{Time-lapse series of nose ventilation captured underwater at a varying temporal resolution (supplementary video clip $1$).
    The flow is from left to right.
    Timestamp within brackets given in seconds and units of convective timescale $t(u/c)$.
    Test case: NACA 0010-34 at $Fr = 1.5$ and $AR = 1$.
    }
    \label{fig: trigger_N}
\end{figure}

Nose ventilation is characteristic of hydrofoils developing a \gls{lsb}, before widespread leading-edge stall \citep{McCullough1951}, at moderate to high angles of attack.
A time-lapse series of this trigger mechanism is given in figure \ref{fig: trigger_N} at a varying temporal resolution  (supplementary video clip $1$).
Immediately before the onset of ventilation, flow visualisations using oil film \citep{Wetzel1957, Breslin1959} and air injection \citep{Wadlin1958} reveal a separated flow region extending along the span without ever reaching the free surface, as illustrated in frame $1$.
The ensuing vortex within this region contributes to locally lowering the static pressure to subatmospheric levels.
However, the interfacial layer of attached flow forming between the separated flow region and the free surface effectively acts as a seal, preventing the establishment of a steady air source.
As the angle of attack increases, this interfacial layer gradually becomes more prone to rupture under sufficiently large surface perturbations.
Once the free surface is breached, air rushes into the separated flow region, developing a ventilated cavity.
The process takes about $0.4\,$s or $3.5\tau$ for the case illustrated in figure \ref{fig: trigger_N} corresponding to the NACA 0010-34 at a Froude number of $1.5$ and an aspect ratio of $1.0$.

\begin{figure}
    \centering
    \includegraphics{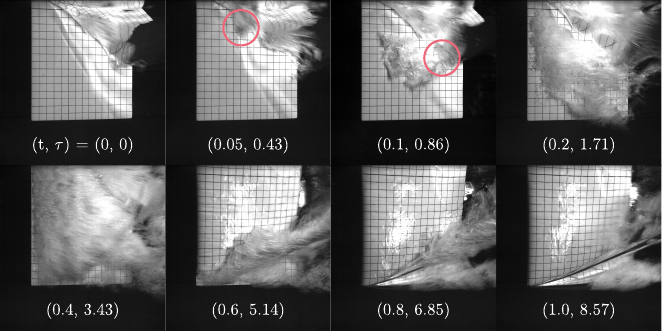}
    \caption{Time-lapse series of tail ventilation captured underwater at a varying temporal resolution (supplementary video clip $2$).
    The flow is from left to right.
    Timestamp within brackets given in seconds and units of convective timescale $t(u/c)$.
    Test case: NACA 0010-34 at $Fr = 1.5$ and $AR = 1$.
    }
    \label{fig: trigger_RT}
\end{figure}

The second trigger mechanism is associated with the \emph{Rayleigh–Taylor} instability of the free surface, subject to the downward acceleration created by the hydrofoil moving through nominally flat water \citep{Taylor1950}.
A time-lapse series is given in figure \ref{fig: trigger_RT} (supplementary video clip $2$).
Immediately before the onset of ventilation, a low-pressure region of detached flow develops over the suction side of the hydrofoil.
Surface perturbations --- either too small or occurring too far downstream from the leading edge to induce nose ventilation --- subject to this low-pressure region develop into air-filled vortices as they travel along the chord.
With increasing angle of attack, the pressure further decreases, and the vortices grow larger, eventually allowing air to enter the separated flow region \citep{Rothblum1969, Swales1974}.
The entire process, from the initial perturbation through vortex formation to the development of a ventilated cavity, takes approximately $0.8\,$s or $7\tau$.
The development time is about twice that of the process illustrated in figure \ref{fig: trigger_N}, a separate run of the same test case which exhibited a different trigger mechanism.
This trend is consistent across the data set, suggesting that tail ventilation is a more gradual process than nose ventilation.

\begin{figure}
    \centering
    \includegraphics{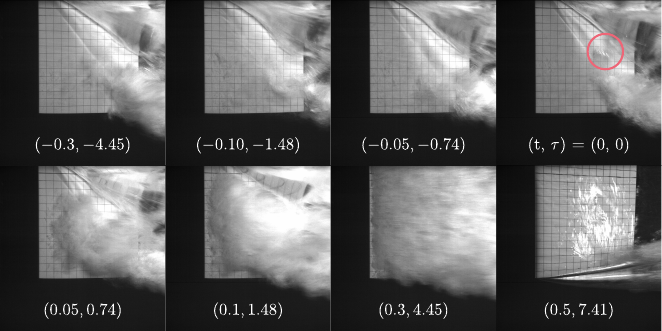}
    \caption{Time-lapse series of base ventilation followed by tail ventilation captured underwater at a varying temporal resolution (supplementary video clip $3$).
    The flow is from left to right.
    Timestamp within brackets given in seconds and units of convective timescale $t(u/c)$.
    The origin is the point at which the air breached the free surface, forming a stable ventilated cavity.
    Test case: semi-ogive at $Fr = 2.5$ and $AR = 1$.
    }
    \label{fig: trigger_RT-N}
\end{figure}

The semi-ogive displayed a third trigger mechanism characteristic of designs with a blunt trailing edge, like wedge-shaped or truncated profiles, referred to as base ventilation \citep{Harwood2014, Harwood2016}.
This trigger mechanism occurred at shallow angles of attack but did not ultimately lead to the formation of a stable ventilated cavity under the examined test conditions.
A time-lapse series is shown in figure \ref{fig: trigger_RT-N} (supplementary video clip $3$).
In the leading period to the onset of ventilation, a wake vortex develops through which air propagates, creating an intermittent air pocket confined to a small region near the trailing edge.
The air pocket is visible over the suction side, extending half the chord length at $t=-0.3\,$s.
It is then partially washed off at  $t=-0.1\,$s and grows larger again at $t=0\,$s.
At this point, a surface perturbation subjected to Rayleigh-Taylor instability develops into an air-filled vortex, growing sufficiently large to allow air to enter the separated flow region.
Tail ventilation is thus the trigger mechanism that ultimately leads to the formation stable ventilated cavity, while base ventilation acts as an underlying precursor mechanism.
The ventilated cavity establishes within $0.5\,$s, or $7\tau$, a period comparable to the test case illustrated in figure \ref{fig: trigger_RT}.

\begin{figure}
    \centering
    \includegraphics{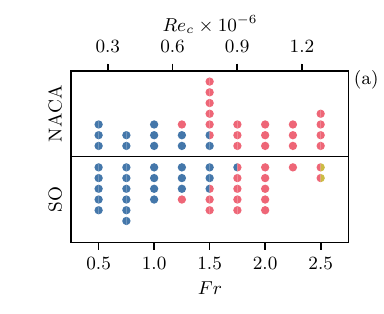}
    \includegraphics{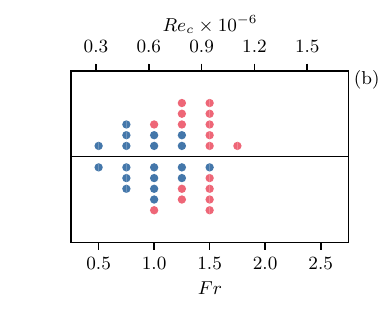}
    \vspace{-10pt}
    \caption{Observed trigger mechanisms for each run as a function of the depth-based Froude number for aspect ratios (a) $AR = 1.0$ and (b) $AR = 1.5$.
    Different coloured markers indicate nose ventilation (blue), Taylor-Rayleigh ventilation (red), and base ventilation (yellow).
    Split-coloured markers indicate two concurrent trigger mechanisms.}
    \label{fig: trigger_mechanisms}
\end{figure}

Tip-vortex ventilation is another trigger mechanism associated with surface-piercing hydrofoils \citep{Ramsen1957, Wetzel1957, Breslin1959, Swales1974} but did not emerge in the present study.
According to \citet{Young2017}, it occurs when the contraction of the wake and potential buoyancy (in the case of cavitation) lift the tip vortex towards the free surface, creating a pathway between the atmosphere and a separated flow region.
Neither the semi-ogive nor the NACA 0010-34 showed signs of tip vortex ventilation.
Still, evidence suggests that base ventilation can effectively bypass the tip vortex by providing a shorter pathway to the tip along the trailing edge \citep{Harwood2016}.
This proxy trigger mechanism was observed for the semi-ogive at a Froude number of $2.5$ and an aspect ratio of $1.0$.
As illustrated in the supplementary video clip $3$, base ventilation induces a mode of ventilation akin to tip vortex ventilation, which does not lead to the formation of a stable ventilated cavity.

Figure \ref{fig: trigger_mechanisms} provides an overview of the trigger mechanisms against the depth-based Froude number and the chord-based Reynolds number\footnote{The specified Reynolds number corresponds to the NACA 0010-34 airfoil, which is approximately $10\,$\% higher than that of the semi-ogive.} along the bottom and top horizontal axes, respectively.
Despite the limited number of runs, the results reveal a clear trend: nose ventilation is the prevalent trigger mechanism at low Froude numbers, whereas tail ventilation is prevalent at moderate to high Froude numbers.
Nose ventilation was observed concurrently with tail ventilation for both model hydrofoils at an aspect ratio of $1.0$ (refer to supplementary video clip $4$).
Base ventilation occurred only for the semi-ogive at an aspect ratio of $1.0$.
It may take over at Froude numbers beyond $2.5$, but there is insufficient evidence to confirm this hypothesis.
The transition between trigger mechanisms occurs gradually. 
For an aspect ratio of $1.0$, tail ventilation does not emerge until a Froude number of $1.25$ and only becomes the sole trigger mechanism at a Froude number of $1.75$.
Similarly, for an aspect ratio of $1.5$, the transition occurs between Froude numbers of $1.0$ and $1.5$.
The equivalent range of Reynolds numbers from $6\times10^5$ to $9 \times 10^5$ aligns with the laminar-to-turbulent transition of the boundary layer.
In agreement with previous studies \citep[e.g.,][]{Breslin1959, Young2017}, these results indicate that as the boundary layer becomes turbulent, the formation of a \gls{lsb} is increasingly unlikely, and, consequently, so is nose ventilation.
A more detailed discussion on the transition between trigger mechanisms is provided in \S \ref{subsec: dynamics_trigger_mechanisms}.

\subsection{Onset of ventilation}
\label{subsec: onset_ventilation}

\begin{figure}
    \centering
    \includegraphics{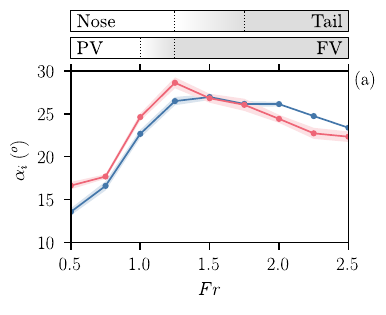}
    \includegraphics{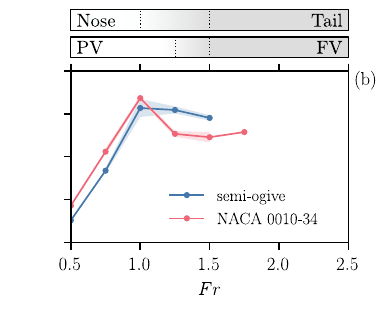}
    \includegraphics{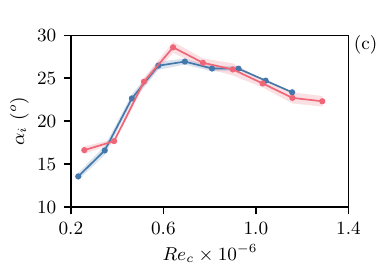}
    \includegraphics{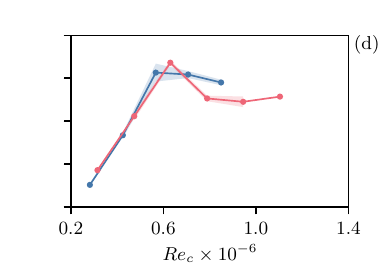}
    \vspace{-10pt}
    \caption{Inception angle of attack $\alpha_i$ as a function of the depth-based Froude number $Fr$ and the chord-based Reynolds number $Re_c$ for aspect ratios $1.0$, (a) and (c), and $1.5$, (b) and (d). 
    The shaded regions around the curves represent the $95\,$\% expanded uncertainty (see appendix \ref{app: A} for further details).
    The top greyscale indicates the prevalent trigger mechanism, while the bottom greyscale indicates the transition between flow regimes, from the \gls{fw} regime to the \gls{pv} or \gls{fv}.}
    \label{fig: alpha_i}
\end{figure}

Ventilation \emph{inception} is defined as the point at which the air breaches the free surface, prompting the formation of a stable ventilated cavity.
As the cavity develops, the flow transitions from the \gls{fw} regime to the \gls{pv} or the \gls{fv} regime.
This definition differs from that of \citet{Harwood2016}, who define ventilation inception as ``[...] the flow transition from the \gls{fw} to the \gls{pv} regime, marking the first stage in the formation of a ventilated cavity''.
Their observations suggest that the flow invariably transitions to the \gls{pv} regime before eventually reaching the \gls{fv} regime through a process referred to as \emph{stabilisation}.
In contrast, the following analysis shows that the flow may transition directly to the \gls{fv} regime, bypassing the \gls{pv} regime altogether.
This mismatch arises from the different approaches to surveying the parameter space, as becomes evident below.

Figure \ref{fig: alpha_i} shows the inception angle of attack $\alpha_i$ as a function of the depth-based Froude number and the chord-based Reynolds number for the semi-ogive and the NACA 0010-34 at two aspect ratios.
Despite having markedly different cross-sections, the models exhibit similar trends.
For an aspect ratio of $1.0$, the inception angle of attack rises from approximately $15^\circ$ at a Froude number of $0.5$ to a peak exceeding $25^\circ$ at $1.25$. 
Further increasing the Froude number beyond this point to $2.5$, the inception angle of attack gradually decreases, yet it remains above $20^\circ$.
A similar trend emerges for an aspect ratio of $1.5$, with the inception angle of attack rising from $15^\circ$ at a Froude number of $0.5$ to a peak exceeding $25^\circ$ at $1.0$. 
The semi-ogive then exhibits a gradual decline, whereas the NACA 0010-34 experiences a sudden drop followed by a plateau. 
The peak value for an aspect ratio of $1.5$ is marginally lower and occurs at a lower Froude number than for an aspect ratio of $1.0$.
Plotting the inception angle of attack against the chord-based Reynolds number shifts the curves horizontally such that the peaks align at $6\times10^5$.
This value corresponds approximately to the laminar-to-turbulent transition of the boundary layer \citep{Breslin1959, Young2017}.

At the onset of ventilation, the flow transitioned spontaneously from the \gls{fw} regime to the \gls{pv} or the \gls{fv} regime.
As evidenced by the lower greyscale in figure \ref{fig: alpha_i}, both models experience a transition to the \gls{pv} regime at Froude numbers below $1.25$ and $1.5$ for aspect ratios of $1.0$ and $1.5$, respectively.
At higher Froude numbers, the flow transitions directly to the \gls{fv} regime.
Unlike the prevalent trigger mechanism, there is no clear relationship between the ventilated regime and the Froude number or the Reynolds number.
Within the narrow range indicated by the colour gradient, the cavity displays characteristics of both ventilated regimes, making classification challenging. 
Still, as the angle of attack increases beyond the inception point, the flow gradually takes on more pronounced characteristics of the \gls{fv} regime.

\begin{figure}
    \centering
    \includegraphics{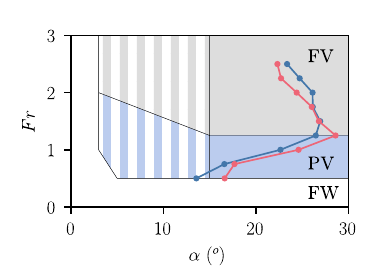}
    \vspace{-10pt}
    \caption{Simplified flow stability map for the semi-ogive hydrofoil at an aspect ratio of $1.0$, adapted from \citet{Harwood2016}, overlaid with the measurements of the inception angle of attack shown in figure \ref{fig: alpha_i}.
    Solid colours represent regions of global stability, and striped patterns represent bistable regions.
    Further details are included in the legend of figure \ref{fig: stability_map_1}.
    }
    \label{fig: stability_map_2}
\end{figure}

Overlaying the measurements of the inception angle of attack onto the flow stability map proposed by \citet{Harwood2016}, as shown in figure \ref{fig: stability_map_2}, reveals a stark discrepancy between the two data sets.
The stability map suggests that, while gradually increasing the angle of attack at a constant Froude number, ventilation should set in at approximately $15\,^{\circ}$, approaching the boundary of the global stability regions of the \gls{pv} and the \gls{fv} regimes.
Instead, ventilation occurs at consistently higher angles of attack than anticipated across nearly the entire range of Froude numbers.
It is delayed compared to the current description of the flow stability map.
This discrepancy is discussed in detail in section \S \ref{subsec: stability_map}, where a revised stability map is proposed to reconcile the two data sets.

\subsection{Hydrodynamic coefficients}
\label{subsec: hydrodynamic_coefficients}
This section examines the hydrodynamic coefficients of the model hydrofoils obtained under steady-state and quasi-steady-state conditions.
The parameters were varied along the $Fr-\text{axis}$ to reach steady-state conditions.
The model hydrofoil was accelerated to a target Froude number at a fixed angle of attack, and the measurements were taken with both parameters held constant.
In contrast, the quasi-steady-state measurements were taken while varying the parameters along the $\alpha-\text{axis}$.
The hydrodynamic coefficients of drag, lift, and the mid-chord yawing moment are given as
\begin{align}
    C_D = - \frac{2 F_x}{\rho u ^2 h c}, \quad
    C_L = \frac{2 F_y}{\rho u ^2 h c}, \quad
    \text{and} \quad
    C_M = \frac{2 M_z}{\rho u ^2 h c^2},
\end{align}
respectively, where $\rho$ is the density of the water.
The time series of the forces and the moment were low-pass filtered at a cut-off frequency determined by the inverse of the convective timescale $1/\tau$.

\subsubsection{Steady state}
\begin{figure}
    \centering
    \includegraphics{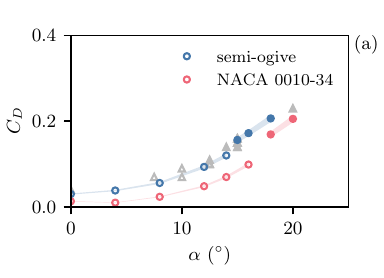}
    \includegraphics{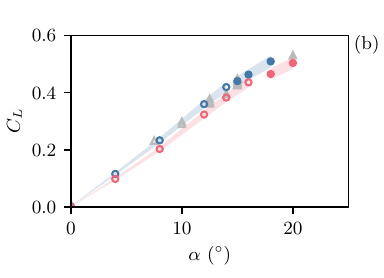}
    \includegraphics{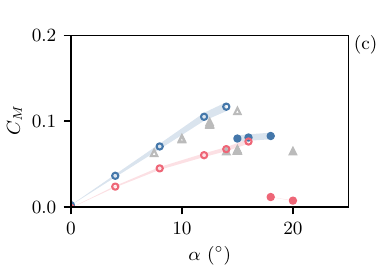}
    \vspace{-10pt}
    \caption{Steady-state hydrodynamic coefficients of drag $C_D$ (a), lift $C_L$ (b), and mid-chord yawing moment $C_M$ (c), as functions of the angle of attack $\alpha$ for both the semi-ogive and the NACA 0010-34, at a Froude number of $1.5$ and an aspect ratio of $1.0$.
    Empty and full markers indicate pre and post-ventilation, respectively.
    The grey-coloured markers are the hydrodynamic coefficients for the semi-ogive, as reported by Harwood (2016).}
    \label{fig: hydrodynamic_coefficients}
\end{figure}

Figure \ref{fig: hydrodynamic_coefficients} shows the steady-state hydrodynamic coefficients as a function of the angle of attack for the semi-ogive and the NACA 0010-34 at a Froude number of $1.5$ and an aspect ratio of $1.0$.
The hydrodynamic coefficients for the semi-ogive are in agreement with those reported by \citet{Harwood2016}, aside from a minor discrepancy in the mid-chord moment coefficient at larger angles of attack.
The flow remains in the \gls{fw} regime for angles of attack below $15\,^{\circ}$ (empty markers).
Beyond this threshold, ventilation occurs before the flow reaches steady-state conditions (full markers).
An angle of attack of $15\,^{\circ}$ corresponds to the boundary between the bistable and globally stable regions for an aspect ratio of $1$, as illustrated in figure \ref{fig: stability_map_2}.
Thus, varying the parameters along the $Fr-\text{axis}$ instead of the $\alpha-\text{axis}$ produces consistent results with previous studies.
The hydrodynamic coefficients for the NACA 0010-34 show similar trends but also reveal significant differences between the two models.
The streamlined cross-sectional profile reduces the drag coefficient, while the lift coefficient is only marginally lower.
Consequently, the lift-to-drag ratio is considerably higher across the entire range of angles of attack, less so post-ventilation.
The mid-chord moment coefficient is approximately half that of the semi-ogive.
The flow remains in the \gls{fw} regime for angles of attack below $16\,^{\circ}$, a slightly higher threshold, in agreement with observations under quasi-steady-state conditions (refer to figure \ref{fig: alpha_i}).

\subsubsection{Quasi-steady state?}
\label{subsec: quasi-steady state?}
\begin{figure}
    \centering
    \includegraphics{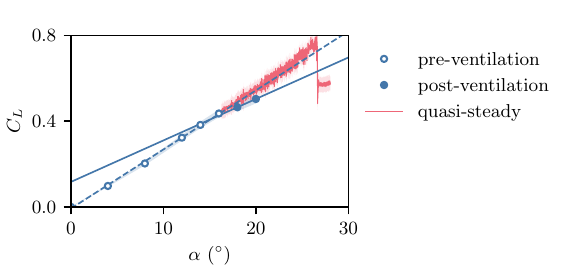}
    \vspace{-10pt}
    \caption{Lift coefficient $C_L$ as a function of the angle of attack $\alpha$ for the model hydrofoil NACA 0010-34 at a Froude number of $1.5$ and an aspect ratio of $1.0$.
    Empty and full markers represent pre and post-ventilation steady-state tests, respectively, and the dashed and solid blue-coloured lines are the linear regressions through those points.
    The solid red-coloured line is the time series of the lift coefficient as the angle of attack increased under quasi-steady-state conditions.
    The time series was low-pass filtered at a cut-off frequency given by the inverse of the convective timescale $1/\tau$.}
    \label{fig: steady_quasi-steady}
\end{figure}

A criterion for the quasi-steady state is derived in \S \ref{subsec: quasi-steady}, based on complex potential flow theory applied to a Joukowski profile undergoing a yawing motion.
It establishes that the lift force is negligibly affected compared to the steady state provided the non-dimensional rate of change of the angle of attack $\dot \alpha^* < 4.36 \times 10^{-4}$ or, equivalently, the number of convective timescales per degree $\Delta\tau > 20$.
To verify whether the measurements meet this criterion, figure \ref{fig: steady_quasi-steady} presents the lift coefficient of the NACA 0010-34 at a Froude number of $1.5$ and an aspect ratio of $1.0$ under both steady-state and quasi-steady-state conditions, as a function of the angle of attack.
The steady-state lift coefficient exhibits a linear trend, represented by the dashed blue-coloured line, and the flow remains in the \gls{fw} regime for angles of attack below $15^{\circ}$.
At higher values, ventilation occurs before reaching steady-state conditions, and the lift coefficient takes a different trend.
The maximum angle of attack at which the hydrofoil could be accelerated without prematurely triggering ventilation set the initial condition for the quasi-steady-state test.
Once the target Froude number was reached, the angle of attack was gradually increased until the onset of ventilation at about $27^\circ$.
The corresponding lift coefficient, represented by the solid red-coloured line, closely follows the pre-ventilation curve before suddenly dropping, confirming that the steady-state assumption is well approximated.

\subsubsection{At the onset of ventilation}

\begin{figure}
    \centering
    \includegraphics{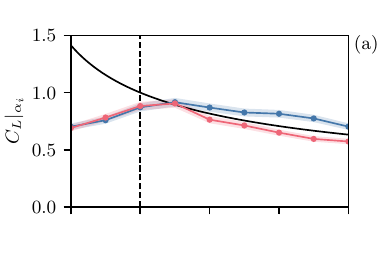}
    \includegraphics{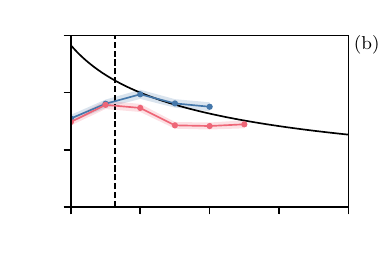}\\[-6mm]
    \includegraphics{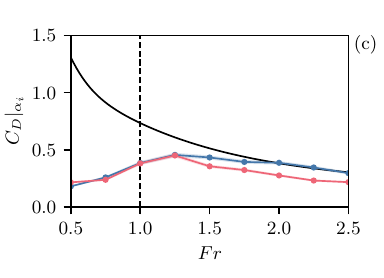}
    \includegraphics{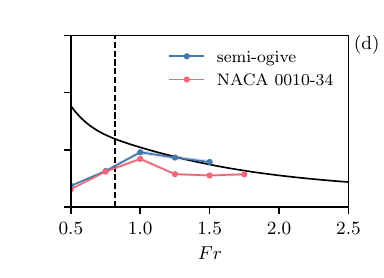}
    \vspace{-10pt}
    \caption{Hydrodynamic lift and drag coefficients, $C_L$ and $C_D$, respectively, at the onset of ventilation against the depth-based Froude number for two aspect ratios.
    The measurements were averaged over a leading period of $10\tau$.
    Solid-black lines are the empirical relationships for lift and drag expressed by equations \ref{eq: empirical_model_lift} and \ref{eq: empirical_model_drag}.
    Dashed-black lines represent the threshold for tail ventilation expressed by equation \ref{eq: RT_condition}.}
    \label{fig: coefficients_model}
\end{figure}

Numerous studies have proposed semi-empirical expressions to characterise the behaviour of surface-piercing hydrofoils based on the flow conditions and, crucially, predict the onset of ventilation.
They have been comprehensively reviewed by \citet{Flatinsen2005}, \citet{Harwood2016}, \citet{Young2017}, \citet{Damley2019}, and \citet{Molland2021}, amonst others.
Notably, \citet{Damley2019} derived through dimensional analysis an expression for the lift coefficient at the onset of ventilation, given by
\begin{equation}
    C_L|_{\alpha_i} = \left[1 - \exp(-\sigma_v Fr) \right] Fr ^ {-1/2},
    \label{eq: empirical_model_lift}
\end{equation}
where
\begin{equation}
    \sigma_v = \frac{2 (p_0 - p_v)}{\rho u ^ 2}
\end{equation}
is the vaporous cavitation number at the free surface, $p_0$ is the atmospheric air pressure, and $p_v \simeq p_0/40$ is the pressure of water vapour at room temperature.
The corresponding drag coefficient accounts for the combined contributions of the frictional, lift-induced, and wave-making resistances.
It is generally expressed as
\begin{equation}
    C_D|_{\alpha_i} = C_D^f(Re_c, t/c) + C_D^i(C_L|_{\alpha_i}, AR) + C_D^w(C_L|_{\alpha_i}, Fr),
    \label{eq: empirical_model_drag}
\end{equation}
with detailed formulations for these terms provided by \citet{Damley2019}.

In this study, the drag and lift coefficients immediately before the onset of ventilation were estimated by averaging the measurements over a leading period of ten convective time scales, $10\tau$. 
The coefficients are plotted in figure \ref{fig: coefficients_model} as functions of the depth-based Froude number for two aspect ratios, alongside the semi-empirical relationships defined in equations \ref{eq: empirical_model_lift} and \ref{eq: empirical_model_drag}.
The threshold for tail ventilation, derived in \S\ref{subsubsec: Tail ventilation}, is also included for reference.
Similarly to the inception angle of attack, the drag and lift coefficients at the onset of ventilation exhibit different trends correlated with the prevalent trigger mechanism.
The trend is positive at low Froude numbers, where nose ventilation is prevalent.
The measurements for the semi-ogive and the NACA 0010-34 closely align with each other but deviate significantly from the model.
Conversely, at moderate to high Froude numbers, tail ventilation is instead the prevalent trigger mechanism and the trend becomes negative.
In this range, the model provides a reasonably accurate prediction, but the agreement between the measurements deteriorates.
\citet{Damley2019} do not explicitly define a domain of application for the semi-empirical relationships, but the results suggest they must be only valid beyond the threshold for tail ventilation.
The lack of agreement between the measurements for the semi-ogive and the NACA 0010-34 at higher Froude numbers suggests that additional factors, such as the cross-sectional profile of the hydrofoil, may also play a role.
The significant scatter of the data from previous studies around the model, illustrated in figure \ref{fig: coefficients_model_log}, further supports this observation.

\section{Discussion}
\label{sec: discussion}
The following discussion covers two important aspects of ventilation of surface-piercing hydrofoils, supported by empirical evidence presented earlier and findings from published studies.
The first focuses on the dynamics of the trigger mechanisms and the second on the stability map of the flow regimes.

\subsection{Dynamics of trigger mechanisms}
\label{subsec: dynamics_trigger_mechanisms}
As demonstrated in \S \ref{subsec: onset_ventilation}, positive and negative trends in the inception angle of attack are associated with distinct trigger mechanisms, namely nose and tail ventilation.
This observation highlights the need for a thorough understanding of the underlying physical processes to explain the overall trend and develop reliable predictive models for ventilation.
Accordingly, the following section attempts to provide an interpretation of the dynamics of these trigger mechanisms.

\subsubsection{Nose ventilation}
The development of a \gls{lsb} with subsequent reattachment, associated with leading-edge stall, has long been identified as a precursor of nose ventilation \citep{Breslin1959, Wadlin1958, Swales1974}.
Its behaviour is governed by the chord-based Reynolds number and the angle of attack.
\citet{McCullough1951} demonstrated that varying the angle of attack at a fixed Reynolds number modifies the pressure distribution around the hydrofoil, which in turn influences the location and extent of the \gls{lsb}.
They argued that increasing the angle of attack leads to a corresponding increase in the adverse pressure gradient behind the leading-edge suction peak.
A stronger adverse pressure gradient promotes a quicker transition of the boundary layer flow from laminar to turbulent, prompting earlier reattachment.
The \gls{lsb} shifts forward, becomes shorter along the chord-wise direction, and the static-pressure coefficient within it decreases.
Although the precise mechanism remains unclear, these effects combined appear to favour nose ventilation.
They presumably contribute to reducing the thickness of the interfacial layer between the free surface and the separated flow region.
In contrast, increasing the Reynolds number at a fixed angle of attack causes the flow to transition and reattach earlier under the same adverse pressure gradient.
The \gls{lsb} becomes progressively smaller while the static-pressure coefficient within it remains unchanged. 
The reduction in the thickness of the interfacial layer is then less pronounced, leading to a delay in nose ventilation.
As shown in figure \ref{fig: alpha_i}, the inception angle of attack $\alpha_i$ increases until the threshold for tail ventilation.

The behaviour of the \gls{lsb} described above is not only consistent with findings from \citet{McCullough1951} but also \citet{Gault1957}, \citet{Roberts1980}, and \citet{Bastedo1986}. 
These studies focus exclusively on two-dimensional, single-phase flows whose topology differs from those around surface-piercing hydrofoils, where factors like free-surface deformation and tip vortex come into play.
Thus, they do not offer a complete description of the relevant physical processes.
Particularly, the interaction between the \gls{lsb} and the free surface leading to nose ventilation remains largely unexplored.
Another significant limitation is that few studies on the ventilation of surface-piercing hydrofoils establish the difference between \emph{leading-edge stall}, preceded by the development of a \gls{lsb}, and \emph{thin-hydrofoil stall}, as formally defined by \citet{McCullough1951}.
The latter is characteristic of hydrofoils with a sharp, wedge-shaped cross-section or, in some cases, a rounded leading edge with a thickness ratio below $9\,\%$. 
The flow around this type of hydrofoil separates at the leading edge and reattaches farther downstream, but, in the absence of a \gls{lsb}, the reattachment point shifts progressively aft as the angle of attack increases.
Consequently, nose ventilation may not emerge, and the inception angle of attack would likely follow a different trend.
Documented examples of thin-hydrofoil stall can be found in \citet{Gault1957} and \citet{Heckel1974}.

\subsubsection{Tail ventilation}
\label{subsubsec: Tail ventilation}
Tail ventilation is driven by the downward acceleration acting on the suction side of the hydrofoil moving through nominally flat water  \citep{Rothblum1969, Swales1974}.
Surface perturbations subject to this acceleration field can become unstable and develop into air-filled vortices as they travel along the chord.
Under suitable conditions, they grow sufficiently long to let air into a separated flow region at subatmospheric pressure.
A semi-empirical model for tail ventilation can be derived based on the theoretical framework for the instability of an interface between two fluids $\zeta$, as outlined by \citet{Taylor1950}.
Neglecting the density of  air relative to the density of water, the amplification factor of a surface perturbation as the ratio to its initial amplitude $\eta/\eta_0$, under a vertical downward acceleration $\ddot \zeta$, at any time $t$, is expressed as
\begin{equation}
    \frac{\eta}{\eta_0} = \cosh \left\{\left[ k \left(\ddot{\zeta} - g\right) \right] ^ {1/2} t\right\},
    \label{eq: eta_amplification_1}
\end{equation}
where $k = 2\pi/\lambda$ is the angular wavenumber and $\lambda$ is the wavelength.

The downward acceleration $\ddot{\zeta}$ created by the hydrofoil is not readily available but can be estimated by assuming it to be uniform along the chord and considering the free-surface deformation at the trailing edge $\Delta \zeta$, illustrated in figure \ref{fig: model}a.
From the equation of motion for a fluid element on the surface, initially at rest and then suddenly set into motion,
\begin{equation}
    \ddot\zeta = \frac{2\Delta\zeta}{\Delta t^2},
\end{equation}
where $\Delta t$ is the time it takes for the fluid element to travel $\Delta\zeta$.
For all test cases exhibiting tail ventilation, measurements show that $\Delta\zeta \simeq c/2$ \citep[][and figure \ref{fig: zeta}]{McGregor1973, Swales1974}.
Introducing this kinematic condition and approximating $\Delta t$ by the convective timescale, the downward acceleration reduces to
\begin{equation}
    \ddot{\zeta} \simeq \frac{u^2}{c}.
    \label{eq: zeta_ddot}
\end{equation}
Replacing equation \ref{eq: zeta_ddot} into equation \ref{eq: eta_amplification_1} and approximating the time that the perturbation is subject to the acceleration field as $t \simeq x/u$, the amplification factor can be expressed as
\begin{equation}
    \frac{\eta}{\eta_0} = \cosh \left\{\left[ k \left(\frac{u^2}{c} - g\right) \right] ^ {1/2} \frac{x}{u}\right\}.
    \label{eq: eta_amplification_2}
\end{equation}

Equation \ref{eq: eta_amplification_2} establishes a necessary condition for tail ventilation.
Specifically, for any surface perturbation to grow unstable, the downward acceleration created by the hydrofoil must exceed the gravitational acceleration \citep{Taylor1950}.
This condition is expressed as $(u^2/c - g) > 0$, from which follows that
\begin{align}
    Fr > AR\,^{-1/2},
    \label{eq: RT_condition}
\end{align}
valid at least for $AR \sim \mathcal{O}(1)$ and $Fr < 2.5$.
Equation \ref{eq: RT_condition} implies that the threshold for tail ventilation is independent of the sea state and inversely proportional to the aspect ratio.
For example, tail ventilation can only occur at Froude numbers higher than $1$ for an aspect ratio of $1$ and higher than $0.82$ for an aspect ratio of $1.5$.
Assuming that surface perturbations are statistically similar across all test conditions, the difference of $0.18$ between these threshold values likely explains the horizontal shift in the peak of the inception angle of attack observed in figure \ref{fig: alpha_i}.

\begin{figure}
    \centering
    \includegraphics{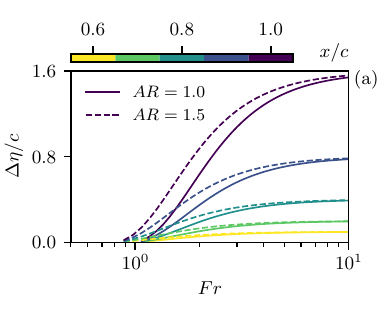}
    \includegraphics{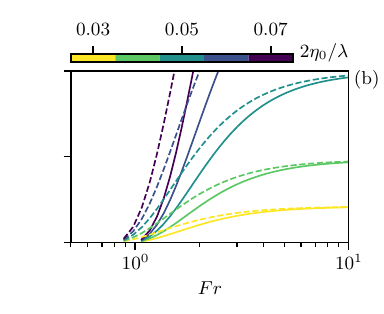}
    \vspace{-10pt}
    \caption{Predicted amplitude increase of a surface perturbation with an initial amplitude $\eta_0 = 1\,\text{mm}$ travelling along the chord as a function of the Froude number for two aspect ratios.
    (a)  Fixed wave steepness ratio $2\eta_0/\lambda = 0.05$ and varying chord-wise locations from mid-chord to the trailing edge in increments of $0.1c$.
    (b)  Fixed chord-wise location $x/c = 1$ and varying wave steepness ratio in increments of $0.01$.}
    \label{fig: taylor}
\end{figure}

When the condition for tail ventilation is satisfied, equation \ref{eq: eta_amplification_2} describes the development of a surface perturbation as it travels along the chord.
Two competing factors influence this process: as the speed of the hydrofoil increases, (1) the downward acceleration field becomes more intense ($u^2/c$), and (2) the residence time of the perturbation within this field decreases ($x/u$).
To illustrate their combined effect, consider an arbitrarily small surface perturbation, with an initial amplitude $\eta_0 = 1\,\text{mm}$ and a wave steepness ratio of $2\,\eta_0/\lambda = 0.05$.
The predicted amplitude increase $\Delta \eta/c$ at evenly spaced chord-wise locations is shown in figure \ref{fig: taylor}a, as a function of the Froude number for aspect ratios $1.0$ and $1.5$.
The growth of the surface perturbation appears to be exponential, nearly doubling the amplitude every $0.1c$, as evidenced by the spread of the contour lines.
The amplification factor at any fixed chord-wise location increases steadily before reaching a plateau.
This behaviour suggests that at low Froude numbers, downward acceleration is the dominant factor, whereas, at high Froude numbers, residence time plays a more significant role.

In the limit when downward acceleration far exceeds gravitational acceleration at high Froude numbers, i.e. $u^2/c \gg g$, equation \ref{eq: eta_amplification_2} implies that the growth rate per unit distance approaches $(k/c)^{1/2}$.
Under this condition, the growth rate depends only on the wavenumber of the perturbation and the chord length, as the above-mentioned competing factors balance each other out.
Assuming the inception angle of attack is inversely proportional to the growth rate, it follows that, after the transition from nose ventilation to tail ventilation, the inception angle of attack should initially decrease and then level off at a lower value --- provided no other trigger mechanisms emerge.
The cause-effect relationship introduced in this argument is strictly not linear, and additional, unaccounted factors associated with the flow topology may also be important.
Nevertheless, the agreement between the prediction and the downward trend observed in figure \ref{fig: alpha_i} suggests that the dynamics of tail ventilation are well captured.

Figure \ref{fig: taylor}b illustrates the effect of the initial wave steepness of a surface perturbation on the predicted amplitude at a fixed chord-wise location as a function of the Froude number.
It highlights a critical challenge in investigating the onset of ventilation of surface-piercing hydrofoils, which is conditional on the characteristics of the free surface.
An increase in wave steepness of $0.01$ more than doubles the amplitude factor.
In this experiment, most surface perturbations originated from small droplets formed by the breakup of a thin water film running up the hydrofoil around the leading edge.
The size of these droplets, as well as the frequency and location of their impact on the surface, were not characterised.
It is assumed that these factors remain statistically consistent across all test conditions.
Had a different mechanism been observed, the inception angle of attack would have likely been modified while preserving the overall downward trend.

\subsection{Revised stability map}
\label{subsec: stability_map}
\begin{figure}
    \centering
    \includegraphics{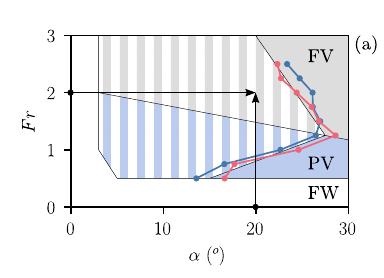}
    \includegraphics{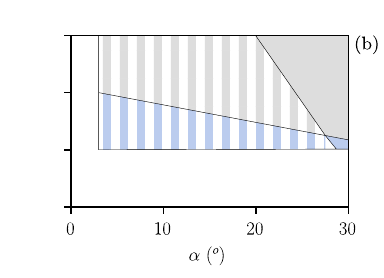}
    \vspace{-10pt}
    \caption{(a) Revised flow stability map obtained by reshaping the boundaries of the global stability regions to match the measurements of the inception angle of attack presented in figure \ref{fig: alpha_i}.
    Further details are included in the legend of figure \ref{fig: stability_map_1}.
    (b) Presumed flow stability map for thick hydrofoils that would be less vulnerable to nose ventilation.
    }
    \label{fig: stability_map_revised}
\end{figure}

In \S \ref{subsec: onset_ventilation}, the inception angle of attack is shown to be significantly higher than predicted by the stability map proposed by \citet{Harwood2016}. 
A similar discrepancy was reported by \citet{Charlou2019}, who conducted numerical simulations (RANS $k$-$\omega$ SST model) of the flow around the semi-ogive hydrofoil.
They surveyed the parameter space along the $Fr-\text{axis}$, by increasing the Froude number at a fixed angle of attack and, conversely, along the $\alpha-\text{axis}$ by increasing the angle of attack and maintaining a constant Froude number.
Their results show that ventilation is delayed by varying the parameters along the $\alpha-\text{axis}$, occurring at angles of attack exceeding $25\,^{\circ}$ at a Froude number of $2.5$.
In contrast, travelling along the $Fr-\text{axis}$, ventilation occurred for angles attack as low as $12\,^{\circ}$.
In these simulations, the linear acceleration of the hydrofoil was twice that reported by \citet{Harwood2016}, and the rate of change of the angle of attack was an order of magnitude higher than in the present study.
\citet{Charlou2019} then concluded that the onset of ventilation is affected by transient phenomena on the path to an arbitrary steady-state condition.

Dynamic motion may undoubtedly affect the onset of ventilation \citep{Young2022, Young2023a, Young2023b} but does not account for the discrepancy observed in this experiment, conducted under quasi-steady-state conditions.
To reconcile the measurements with those reported by \citet{Harwood2016}, a revised stability map is proposed in figure \ref{fig: stability_map_revised}a that significantly extends the bistable flow regions.
According to the revised stability map, two alternative paths to a steady-state condition can lead to different flow regimes.
For example, varying the parameters along the $Fr-\text{axis}$ for angles of attack above $15\,^{\circ}$, the flow is expected to transition from the \gls{fw} to the \gls{pv} regime at a Froude number of $0.5$.
As the Froude number continues to rise, the hydrofoil enters a bistable flow region where the \gls{pv} regime is locally stable.
The flow does not revert to the \gls{fw} regime but remains ventilated.
At even higher Froude numbers, the ventilated cavity grows larger, and the flow naturally transitions to the \gls{fv} regime, crossing over to the \gls{fw}/\gls{fv} bistable region.
Alternatively, travelling along the $\alpha-\text{axis}$, the flow remains in the \gls{fw} regime unless a sufficiently large surface perturbation induces a transition to a locally stable ventilated regime.

Far from universal, the proposed stability map depends not only on the cross-sectional profile of the hydrofoil and the aspect ratio but also on additional geometrical parameters, such as sweep and roll angles \citep{Young2017, Matveev2018, Huang2024}.
Deriving a stability map for every possible parameter combination is virtually impossible. 
However, a deeper understanding of the underlying physical mechanisms can provide valuable insights into how these factors influence stability.
In particular, the following section presents a series of arguments to extend the stability map to hydrofoils with thick cross-sectional profiles.

\subsubsection{Variation for thick hydrofoils}
Nose ventilation is characteristic of thin hydrofoils developing a \gls{lsb} with subsequent reattachment \citep{Breslin1959, Wadlin1958, Swales1974}.
Thick hydrofoils, experiencing a more gradual pressure recovery, are less vulnerable to this trigger mechanism, which is unlikely to emerge or become prevalent.
It is thus reasonable to assume that the stability map would differ from that schematically illustrated in figure \ref{fig: stability_map_revised}a.
As indicated in figure \ref{fig: stability_map_revised}b, in the absence of nose ventilation, the \gls{fw}/\gls{pv} bistable region (white and blue striped area) would be significantly reduced.
The global stability region of the \gls{pv} regime would also be reduced or cease to exist for aspect ratios lower than $1$, in which case, surveying the parameter space along the $Fr-\text{axis}$ or $\alpha-\text{axis}$ would yield the same outcome.

The present study cannot unequivocally confirm this hypothesis, because the models examined are both relatively thin and prone to nose ventilation.
Yet, it offers three pieces of supporting evidence.
The first relates to the theoretical framework outlined in \S \ref{subsec: dynamics_trigger_mechanisms}, which establishes a necessary condition for the onset of tail ventilation, i.e. $Fr > AR\,^{-1/2}$.
This condition implies that, for thick hydrofoils with an aspect ratio of $1$, the boundary of the \gls{fw}/\gls{pv} bistable region lies at a Froude number of at least $1$ instead of $0.5$ according to the reference stability map.
Furthermore, even higher values would be required for hydrofoils with lower aspect ratios, thereby shrinking the global stability region of the \gls{pv} regime and potentially eliminating it.

The second piece of evidence suggests that above the threshold for tail ventilation and below the global stability region of the \gls{fv} regime, the inception angle of attack would be higher in the absence of nose ventilation.
As illustrated in figure \ref{fig: taylor}a, within this range, the amplification factor of a surface perturbation travelling along the chord increases sharply with the Froude number.
This increase is particularly limited just above the threshold for tail ventilation, which would lead to higher inception angles of attack.
A similar argument can be made by examining the drag and lift hydrodynamic coefficients at the onset of ventilation given in figure \ref{fig: coefficients_model}.
Assuming the semi-empirical model is valid \citep{Damley2019}, tail ventilation would occur only when the hydrodynamic coefficients reach much higher values, corresponding to greater angles of attack.

\begin{figure}
    \centering
    \includegraphics{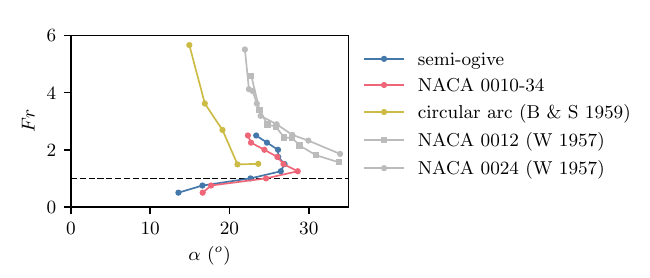}
    \vspace{-10pt}
    \caption{Boundary of the global stability region for the \gls{pv} or \gls{fv} flow regimes for various model hydrofoils with an aspect ratio of $1$ (ventilation occurs to the right-hand-side of each curve).
    The circular arc has a chord length of $2.5\,\text{in}$ \citep{Breslin1959}, while the NACA 0012 and NACA 0024 have a chord length of $2\,\text{in}$ \citep{Wetzel1957}.}
    \label{fig: stability_map_literature}
\end{figure}

Finally, a third piece of evidence is presented in figure \ref{fig: stability_map_literature}, comparing the present measurements with data from earlier studies on the onset of ventilation of relatively thick hydrofoils with an aspect ratio of $1$. \citet{Breslin1959} tested a circular-arc hydrofoil with a $2.5\,\text{in}$ ($63.5\,\text{mm}$) chord length and a thickness ratio of $12\,\%$, and \citet{Wetzel1957} tested two symmetric NACA-profile hydrofoils with chord lengths of $2\,\text{in}$ ($50.8\,\text{mm}$) and thickness ratios of $12\,\%$ and $24\,\%$.
The curves do not collapse for potentially numerous reasons, including differences in the cross-sectional profiles and the susceptibility of the trigger mechanisms to the experimental conditions.
In particular, as illustrated in figure \ref{fig: taylor}b, the ultimate size of surface perturbations subjected to Rayleigh-Taylor instability depends on their initial amplitude, making the onset of ventilation conditional on the characteristics of the free surface and influenced by the presence of surfactants and floating debris.
The trends are nonetheless consistent across data sets.
Similarly to \citet{Harwood2016}, \citet{Breslin1959} and \citet{Wetzel1957} surveyed the parameter space along the $Fr-\text{axis}$, but the flow did not ventilate early on at a Froude number of $0.5$.
Descriptions of the ventilated cavity formation in these early studies suggest that tail ventilation and tip-vortex ventilation (at higher Froude numbers) were the prevalent trigger mechanisms.
Had nose ventilation instead been observed, the boundaries would presumably be more closely aligned with the stability map proposed by \citet{Harwood2016}.

\section{Conclusion}
\label{sec: conclusion}
\glsresetall
Experiments were conducted to investigate the onset of ventilation of surface-piercing hydrofoils.
Under steady-state conditions, the depth-based Froude number and the angle of attack define regions where distinct flow regimes, \gls{fw}, \gls{pv}, and \gls{fv}, are either locally or globally stable. 
The conventional approach for mapping the boundaries between these stability regions consists of systematically surveying the parameter space by setting the model hydrofoil at a fixed angle of attack and increasing the Froude number.
Alternatively, this study surveyed the parameter space by gradually increasing the angle of attack while keeping the Froude number constant.
The rate of change of the angle of attack was sufficiently slow to ensure quasi-steady-state conditions, minimising surface disturbances and the effect of unsteady hydrodynamics.
This criterion was prescribed based on potential flow theory applied to a generic Joukowski profile undergoing a yawing moment, as outlined by \citet{Tupper1983}.
For quasi-steady state conditions, the maximum rate of change of the angle of attack was such that the number of convective timescales per degree $\Delta\tau > 20$.
Two simplified model hydrofoils were examined: the semi-ogive, with a blunt trailing edge, and the NACA 0010-34.
Tests were performed at aspect ratios of $1.0$ and $1.5$ and depth-based Froude numbers ranging from $0.5$ up to $2.5$ in increments of $0.25$, limited by the full scale of the force balance and the loading capacity of the hexapod, which controlled the yaw motion of the model hydrofoils.

For all test conditions, the flow transitioned directly from the \gls{fw} regime to the \gls{pv} or the \gls{fv} regime as the angle of attack gradually increased at a constant Froude number.
Three distinct trigger mechanisms, nose, tail, and base ventilation, were identified depending on the flow conditions.
They are associated with different trends in the inception angle of attack.
Nose ventilation is prevalent at low Froude numbers, below $1.25$ for an aspect ratio of $1.0$ and below $1.0$ for an aspect ratio of $1.5$.
Once the air breaches the free surface, the typical timescale to establish a ventilated cavity is approximately $3.5\tau$.
The inception angle of attack exhibits a positive trend, increasing from $15\,^\circ$ to a peak exceeding $25\,^\circ$.
Tail ventilation is prevalent at higher Froude numbers.
It is a much more gradual process than nose ventilation, with a typical timescale of about $7\tau$.
The inception angle of attack exhibits a negative trend but remains above $20\,^\circ$.
Base ventilation was observed only for the semi-ogive but did not ultimately lead to the formation of a stable ventilated cavity.

To explain the overall trend in the inception angle of attack plotted against the Froude number, the dynamics of the trigger mechanisms of ventilation were examined.
Previously published work established that nose ventilation is associated with the development of a \gls{lsb}, preceding leading-edge stall \citep{McCullough1951, Breslin1959, Swales1974}.
It is hypothesised that the \gls{lsb}, influenced by the Reynolds number and the angle of attack, controls the thickness of the interfacial layer (between the free surface and the separated flow region) and, thereby, the inception angle of attack.
There remains, however, a lack of detailed measurements to elucidate this interaction.
Tail ventilation is a fundamentally different trigger mechanism.
It is driven by the downward acceleration field created by the hydrofoil moving through nominally flat water.
This acceleration field causes surface perturbations to become unstable and develop into air-filled vortices.
A semi-empirical model was derived based on the theoretical framework for the instability of an interface between two fluids, as outlined by \citet{Taylor1950}.
Notably, the model establishes that $Fr>AR^{-1/2}$ is a necessary condition for tail ventilation, valid at least for $AR \sim \mathcal{O}(1)$ and $Fr < 2.5$.
It also indicates that the growth rate of a surface perturbation increases with the Froude number, which translates to a decrease in the inception angle of attack, in agreement with the measurements.

Surveying the parameter space by gradually increasing the angle of attack and maintaining a constant Froude number reveals a stability map which differs significantly from that obtained following the conventional approach \citep{Harwood2016}.
The measurements convincingly show that the boundary between the bistable and globally stable regions, previously estimated to be around $15\,^\circ$ for the semi-ogive, is an artefact of experiment design.
This boundary is not uniform, and, except at a Froude number of $0.5$, it extends to significantly higher angles of attack.
A revised stability map was proposed to reconcile published and present data, elucidating how two alternative paths to a steady state could lead to distinct flow regimes. 
While this revised stability map applies to the semi-ogive and the NACA 0010-34, it is likely not universal. 
Evidence suggests that a thicker hydrofoil, less susceptible or immune to nose ventilation, can achieve higher Froude numbers and angles of attack while remaining in the \gls{fw} regime.

\backsection[Supplementary data]{Supplementary material and movies are available at http://doi.org/[doi].}

\backsection[Acknowledgements]{The authors express their gratitude to the technicians at the Ship Hydromechanics Laboratory of Delft University of Technology for their assistance in successfully completing the experiments.
They also thank Dr. Casey Harwood, Associate Professor at the University of Iowa, for kindly providing access to his dataset upon request.}

\backsection[Funding]{This study was supported through a joint agreement for hydrodynamic research by the Dutch Olympic Committee*Dutch Sports Federation (NOC*NSF), the Royal Netherlands Watersport Association, and Delft University of Technology. 
}

\backsection[Declaration of interests]{The authors report no conflict of interest.}

\backsection[Data availability statement]{The data that support the findings of this study are openly available in [repository name] at http://doi.org/[doi], reference number [reference number].}

\backsection[Author ORCIDs]{M. Aguiar Ferreira, https://orcid.org/0000-0002-2428-0284; C. Navas Rodríguez, https://orcid.org/0009-0002-6584-9200; G. Jacobi, https://orcid.org/0000-0002-9972-3887; D. Fiscaletti, https://orcid.org/0000-0002-3382-3279; A. Greidanus, https://orcid.org/0000-0002-5842-9753; and J. Westerweel, https://orcid.org/0000-0002-2799-5242.}

\backsection[Author contributions]{M. Aguiar Ferreira (ABCDEF), C. Navas Rodríguez (ABCDF), G. Jacobi (ABF), D. Fiscaletti (ABF), A. Greidanus (AFG), and J. Westerweel (AFG) contributed as follows: A – research planning, B – experiment design, C – execution, D – data analysis, E – manuscript preparation, F – editing, and G – funding.}

\appendix
\renewcommand\thefigure{\thesection.\arabic{figure}} 
\setcounter{figure}{0}

\section{Uncertainty analysis}\label{app: A}

The uncertainty analysis follows the guide to the expression of uncertainty in measurement from the International Bureau of Weights and Measures \citep{BIPM2008}. 
Accordingly, the combined uncertainty of a measurement is calculated as the root sum of squares of independent \emph{type A} and \emph{type B} uncertainty contributions,
\begin{equation}  
    u_c = (u_A^2 + u_B^2) ^ {1/2}. 
\end{equation}  
The uncertainty contributions of type A are associated with random errors and are derived through statistical analysis of the dataset, as detailed in \S\ref{subsec: type_A}.
In contrast, uncertainty contributions of type B,  summarised in table \ref{tab: uncertainty}, are associated with systematic errors. 
They are determined based on the specifications of the measurement instrument reported by the manufacturer or calibration.
Throughout the manuscript, all measurements are accompanied by the corresponding expanded uncertainty
\begin{equation}
    u_e = k u_c,
\end{equation}
where the coverage factor $k = 2$ provides a level of confidence of $95\,\%$ associated with the interval defined by $\pm\,u_e$, assuming a normal probability distribution.

\begin{table}
    \begin{center}
        \def~{\hphantom{0}}
        \def\arraystretch{1.2}
        \begin{tabular}{c c c c c}
            $x_i$       & Instrument                        & $\Delta(x_i)$         & $u(x_i)$      & unit\\ \hline
            $c$         & Ruler                             & $1$                   & $0.5$         & mm\\
            $h$         & Datum depth                       & $1$                   & $1$           & mm\\
            $T$         & Thermocouple type K               & $0.1$                 & $1.12$        & $^{\circ}C$\\
            $p_0$       & Setra 278                         & $7.3$                 & $30$          & $\text{Pa}$ \\
            $\rho$      & \S \ref{subsec: water_properties} & -                     & $0.0018\,\%$  & $\text{kg.m}^{-3}$\\
            $\mu$       & ''                                & -                     & $2.6\,\%$     & $\text{Pa.s}$\\
            $\nu$       & ''                                & -                     & $2.6\,\%$     & $\text{m}^2\text{.s}^{-1}$\\
            $u$         & Encoder                           &                       & $1\,\%$       & m.s$^{-1}$\\
            $\alpha$    & \S \ref{subsec: alignment}        & -                     & $0.1$         & $^{\circ}$\\
            $F_x$       & \S \ref{subsec: force_balance}    & $6.1\times10^{-3}$    & $0.1\,\%$     & N\\ 
            $F_y$       & ''                                & $24.4\times10^{-3}$   & $0.1\,\%$     & N\\ 
            $M_z$       & ''                                & $4.9\times10^{-3}$    & $0.5\,\%$     & N.m\\ 
        \end{tabular}
        \caption{Resolution of the measurement of the input quantities and associated type B uncertainty.
        Values specified by the sensor manufacturer or estimated from calibration.}
        \label{tab: uncertainty}
    \end{center}
\end{table}

\subsection{Type A evaluation of uncertainty}
\label{subsec: type_A}
1. Consider an arbitrary quantity $q$ obtained from $n$ independent observations $q_k$ taken under similar conditions.
The arithmetic mean or average is defined as
\begin{equation}
    \overline{q} = \frac{1}{n}\sum_{k=1}^n q_k.
\end{equation}
The experimental standard deviation of the observations $s(q_k)$, characterising the dispersion of the data around the mean, is given by the positive square root of the experimental variance of the observations
\begin{equation}
    s^2(q_k) = \frac{1}{n-1}\sum_{j=1}^n (q_j - \overline{q})^2.
\end{equation}
The experimental variance of the mean
\begin{equation}
    s^2(\overline{q}) =  \frac{s^2(q_k)}{n}.
    \label{eq: variance_mean}
\end{equation}
Thus, the estimate of an input quantity $X_i$ from $n$ independent observations $X_{i,k}$ is expressed as $x_i = \overline{X_i}$ and the corresponding standard uncertainty $u(x_i) = s(\overline{X}_i)$, where $s^2(\overline{X}_i)$ is defined in equation \ref{eq: variance_mean}.

\bigskip
\noindent{
2. The pooled variance $s^2_p$ combines the variances from $m$ independent groups to estimate a common population variance.
It is defined as
\begin{equation}
    s^2_p = \dfrac{\sum\limits_{i=1}^{m} (n_i - 1) s^2_i}{\sum\limits_{i=1}^{m} (n_i - 1)}
    \label{eq: pooled_variance}
\end{equation}
where $s^2_i$ is the experimental variance of the $i$th series of $n_i$ independent repeated observations.
The experimental variance $s^2_p / m$ (and the experimental standard deviation $s_p / m^{1/2}$) of the arithmetic mean of $m$ independent observations characterised by the pooled estimate of variance $s^2_p$ has $\sum_{i=1}^{m} (n_i - 1)$ degrees of freedom.
Particularly, the pooled variance was used to estimate the uncertainty in the measurement of the hydrodynamic forces and moment immediately before the onset of ventilation.
}

\bigskip 
\noindent{
3. The effective number of independent observations for correlated data, such as the time series of the hydrodynamic forces and moment, is estimated by
\begin{equation}
    n^* \, = \, n \frac{\Delta t}{T_e},
\end{equation}
where $\Delta t$ is the sampling period of the signal and $T_e$ is the $e$-folding decay time of autocorrelation (i.e. the point at which the autocorrelation drops below $1/e$).
}

\bigskip 
\noindent{
4. The standard uncertainty of an arbitrary output variable $y = f(x_i)$ is expressed as the positive square root of the combined variance
\begin{equation}
    u^2(y) \, = \, \sum_{i=1}^N u_{x_i}^2(y),
    \label{eq: uncertainty}
\end{equation}
where 
\begin{equation}
    u_{x_i}^2(y) \, = \, \left( \frac{\partial f}{\partial x_i} \right) ^ 2 u^2(x_i)
\end{equation}
and $u^2(x_i)$ is the estimated variance of each independent quantity.
}

\subsection{Water properties}
\label{subsec: water_properties}
The uncertainty in the estimate of the water temperature listed in table \ref{tab: uncertainty} accounts for the $0.5 \,^\circ\mathrm{C}$ measurement error of the type K thermocouple and temperature fluctuations of approximately $1 \,^\circ\mathrm{C}$ observed throughout the measurement campaign. Assuming a negligible effect of atmospheric pressure, the density of the water as a function of temperature is approximated by the quadratic relationship
\begin{equation}
    \rho(T) \, = \, \rho_0 [1 - A (T - T_0) ^ 2],
\end{equation}
where $\rho_0 = 998\,\text{kg.m}^{-3}$ is the reference water density at the reference temperature $T_0 = 20\,^{\circ}C$ and $A = 2.0 \times 10^{-6}\,^{\circ}C^{-2}$ is an empirical coefficient valid for small deviations from the reference condition.
From equation \ref{eq: uncertainty}, it follows that the variance of the water density is
\begin{equation}
    u^2(\rho) \, = \, \left[2 \rho_0 A (T - T_0) \right] ^ 2 u ^ 2(T).
\end{equation}
The dynamic viscosity of water is approximated as
\begin{equation}
    \mu(T) \, = \, B 10\,^{C/ (D + T)},
\end{equation}
where $B = 2.414 \times 10^{-5}\,\text{Pa.s}$, $C = 247.8\,\text{K}$, and $D = 140\,\text{K}$ are empirically determined coefficients, and the corresponding variance
\begin{equation}
    u^2(\mu) \, = \, \left[\frac{B C}{(D + T) ^ 2} \ln(10) 10\,^{C/ (D + T)}\right]^2 u^2(T).
\end{equation}
Finally, the kinematic viscosity of water is defined as
\begin{equation}
    \nu = \frac{\mu}{\rho}
\end{equation}
and the variance of the estimate
\begin{equation}
    u^2(\nu) \, = \, \left(\frac{1}{\rho}\right)^2 u^2(\mu) + \left(\frac{\mu}{\rho^2}\right)^2 u^2(\rho).
\end{equation}

\section{Supplementary figures}
\label{sec: supplementary_figures}
\begin{figure}
    \centering
    \includegraphics{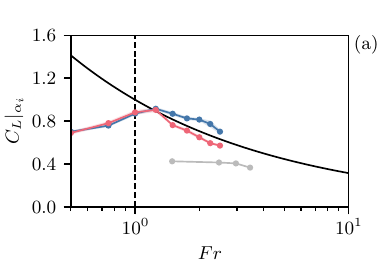}
    \includegraphics{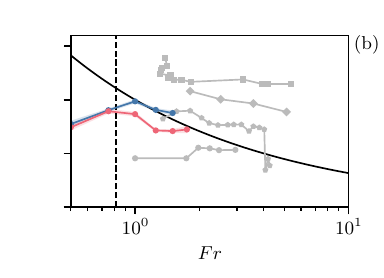}
    \vspace{-10pt}
    \caption{Hydrodynamic lift coefficient at the onset of ventilation against the depth-based Froude number for aspect ratios (a) $AR = 1.0$ and (b) $AR = 1.5$.
    The measurements were averaged over a leading period of $10\tau$.
    Published data from \citet{Harwood2016} $\square$, \citet{Breslin1959} $\circ$, and \citet{Swales1974} $\lozenge$ are included for comparison.
    Solid-black lines are the empirical relationships for lift and drag expressed by equations \ref{eq: empirical_model_lift} and \ref{eq: empirical_model_drag}.
    Dash-black lines represent the threshold for tail ventilation expressed by equation \ref{eq: RT_condition}.}
    \label{fig: coefficients_model_log}
\end{figure}

\begin{figure}
    \centering
    \includegraphics{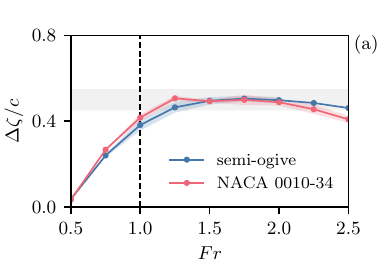}
    \includegraphics{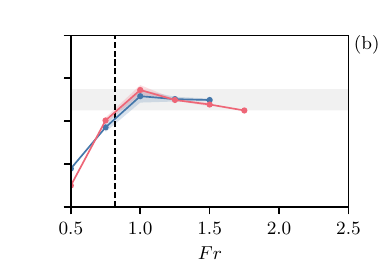}
    \vspace{-10pt}
    \caption{Normalised free-surface deformation at the trailing edge $\Delta \zeta$, defined in figure \ref{fig: model}, as a function of the depth-based Froude number for aspect ratios (a) $AR = 1.0$ and (b) $AR = 1.5$.
    The measurements were averaged over a leading period of $10\tau$.
    Dashed-black lines represent the threshold for tail ventilation expressed by equation \ref{eq: RT_condition}.
    Grey-shaded area is the range $0.45<-\Delta\zeta/c>0.55$.}
    \label{fig: zeta}
\end{figure}

Supplementary figures to provide further clarification on the results discussed in \S \ref{subsec: quasi-steady state?} and \S\ref{subsubsec: Tail ventilation}:
Figure \ref{fig: coefficients_model_log} shows the hydrodynamic lift coefficient at the onset of ventilation as a function of the depth-based Froude number, alongside previously published data from \citet{Breslin1959}, \citet{Swales1974}, and \citet{Harwood2014}.
There is significant scatter around the model proposed by \citet{Damley2019}, suggesting that additional unaccounted factors, such as variations in the cross-sectional profile or the measurement procedure, may play a role.
Figure \ref{fig: zeta} presents the free-surface deformation at the trailing edge normalised by the chord length as a function of the depth-based Froude number.
In the range where tail ventilation is the prevalent trigger mechanisms, Froude numbers exceeding $1.25$ and $1.0$ for aspect ratios $1.0$ and $1.5$, respectively, the free surface is consistently drawn down by approximately $c/2$.
The variation across the range of Froude numbers is within $\pm0.05c$.
This observation is in agreement with those from \citet{McGregor1973} and \citet{Swales1974}.

\bibliographystyle{jfm}
\bibliography{jfm}

\end{document}